# Modeling and Modifying Response of Biochemical Processes for Biocomputing and Biosensing Signal Processing

Sergii Domanskyi and Vladimir Privman*

*Department of Physics, Clarkson University, Potsdam, NY 13676*

**Abstract**

Processes involving multi-input multi-step reaction cascades are used in developing novel biosensing, biocomputing, and decision making systems. In various applications different changes in responses of the constituent processing steps (reactions) in a cascade are desirable in order to allow control of the system's response. Here we consider conversion of convex response to sigmoid by "intensity filtering," as well as "threshold filtering," and we offer a general overview of this field of research. Specifically, we survey rate equation modelling that has been used for enzymatic reactions. This allows us to design modified biochemical processes as "network components" with responses desirable in applications.

* Corresponding author e-mail: privman@clarkson.edu

## 1. Introduction

In theoretical rate-equation modeling of chemical and biochemical reactions in several-step cascades that are being investigated for novel biosensing or biomolecular computing applications,[1,8,9,11,16,17,27,28,40,44,47-49,51,54,55,57,59,62,63,66,78,79,90,105,106,108,117,126,127] one frequently focuses on the select few primary kinetic pathways[53,85,86,99] for each step (reaction, process). This is done in order to limit the number of adjustable parameters in such systems, for which experimental data are typically noisy[72,83] and not sufficiently detailed for a more accurate multi-parameter description of all the possible reaction pathways. Here we illustrate this approach by considering two specific recently studied systems[30,88] relevant to biosensing and biocomputing.[1,17,51,65,107] However, the illustrated general framework for setting up rate-equation modeling applies to many other chemical, biochemical and biomolecular systems in the biosensing and biomolecular computing (biocomputing) contexts, extensively researched over the past decade.[54,99]

We concentrate on processes with multi-input reaction cascades that are used in biosensing, biocomputing, and decision making devices and setups utilizing (bio)chemical processes with well-defined responses.[8,9,49,52,55,57,59,105,106,117,126,127] Enzymatic processes are of particular interest because they promise short-term development of new biosensing[21,40,43,44,59,117,120] and bioactuating applications[58,64,84] with several signal processing steps. Indeed, most biosensing and bioanalytical devices involve enzymatic reactions, which are biocompatible, selective (specific), and also relatively easy to integrate with electronics.[125] For instance, enzyme-based logic systems[8,56,57,65,105,106] operating as binary YES/NO biosensors can be interfaced with electrochemical/electronic devices by coupling to electrodes[43] or field-effect transistors.[55,60,82]

The considered rate equation modelling has been used for cascades of enzymatic reactions.[30,88,90,92] The set of kinetic rate equations describing the key (bio)chemical reaction steps of interconversion of different chemicals as well as the output buildup, is typically solved numerically with finite difference methods. These rate equations model the main reaction steps and enable fitting key process parameters to the extent allowed by limited and/or noisy experimental data. Indeed, the full kinetic description of each enzymatic process would in most cases require numerous parameters (rate constants) for each enzyme. We have developed models[30,88,90,92] that give a reasonable system's response control—and description for potential modifications for applications—with a small number of adjustable parameters.

The use of the rate-equation modeling reviewed here, allows us to "design" modified biochemical processes as "device components" (signal processing steps) with responses desirable in applications. This is illustrated in Figure 1, where panel (a) shows a typical "convex" response of a (bio)chemical process. The output is limited by the input chemical for small inputs, which results in a linear dependence. However, as the input is increased, other chemicals' availability limits the output, and its response to large input values reaches saturation. This can be modified for various applications, as sketched in Figure 1.

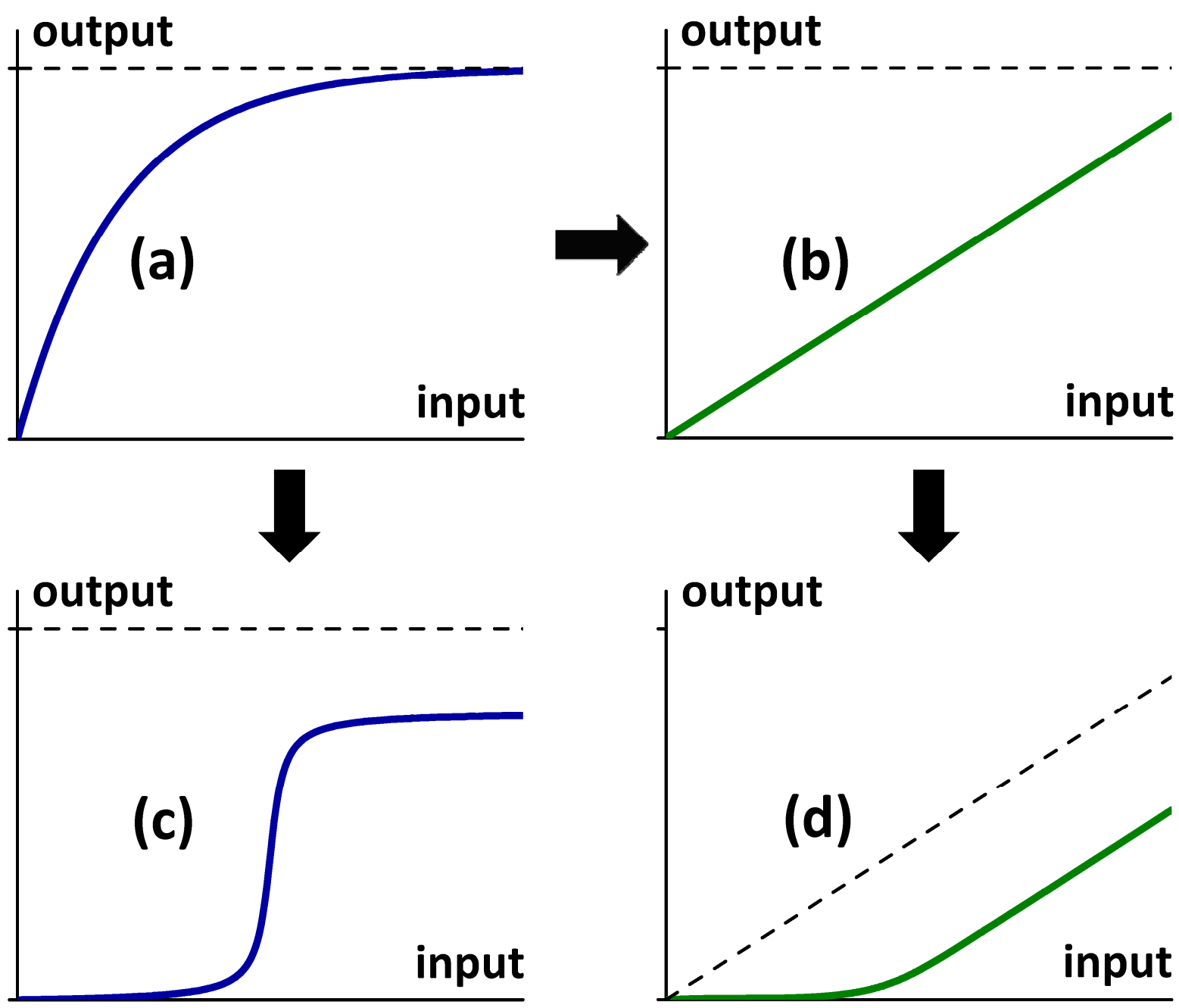

**Figure 1.** (a) A typical "convex" response shape for a chemical or biochemical process. (b) Linear response desirable in many biosensing applications. (c) "Binary" sigmoid-shape response of interest in biomolecular computing, desired to be symmetrical and steep at the middle inflection. (d) In certain applications the conversion of a linear response to the threshold one, followed by a linear behavior, (b) ➡ (d), is required. Adapted with permission from Ref. 88. Copyright © 2014, American Chemical Society.

In Section 2, we offer an illustration of a system where the convex response is modified to yield a sigmoid "filter" shape, by intensity-filtering[6,30,41,42,50,80,89,92,93,95,114,122-124] whereby the input[95] or output[2,42,80] is chemically depleted up to a limited quantity. For such filtering, the dashed line, Figure 1, panel (c), illustrates the possibility of signal loss as the price paid for modifying the system's response. Linear response is desirable in many biosensing applications,[5,20,22,25,29,34,75,87,96,101-103,111,122,128] see Figure 1, panel (b). However, in certain cases threshold response is preferred,[88] as shown in Figure 1, panel (d). In Section 3, we offer an example of a system where an added enzymatic process accomplishes such a response-modification by an interesting new enzyme-functioning mechanism. Section 4 offers a Summary.

## 2. Sigmoid Response and Its Numerical Rate-Equation Modeling

### 2.1. *Sigmoid Response for Noise Reduction in Binary Gate Functioning*

Sigmoid response is useful when "binary" input and output values are of interest in processing based on biomolecular reactions, which has recently been investigated for "digital" sensor and actuator design, logic systems, and other novel ideas in interfacing and computing involving biomolecules.[51,52] Enzyme-catalyzed reactions have been used in such systems, with emphasis on novel diagnostic applications.[4,19,28,51,52,91] For example, binary signal differentiation can be useful for future biomedical and diagnostic applications involving analysis of biomarkers indicative of specific illnesses or injury.[42,44,73,93] Processing steps then mimic binary logic gates and their networks. These developments promise new functionalities for analytical purposes, offering a new class of biosensors which can generate a binary output of the alert type: YES/NO, in response to several input signals. These are parts of biosensor-bioactuator "Sense/Act/Treat" systems.[53,70,115,117] The approach has already been used to analyze biomarkers indicative of certain traumas.[68,81] Binary (digital) in such applications refers to the ability to identify specific values or ranges of values corresponding to **1** or **0** (YES/NO, Act/Don't Act) signals.[86] Standard binary logic gates, including AND, OR, XOR, INHIB, etc.,[2,7,32,91,104,108] and also few-gate model biomolecular networks[3,21,43,90,121] were demonstrated, some mimicking simple digital electronics designs.

Control of noise in functioning of biomolecular gates used as network elements is an important topic to consider.[3,86,91] An effective approach to noise control has been to modify some of the biomolecular reaction responses in a network of processing steps, according to (a) ➡ (c), per Figure 1, i.e., have the output a sigmoid function of the input(s). This mechanism is also used in natural systems.[14,91,98] Sigmoid response then "filters" the output towards the two reference binary values. Such biomolecular filtering based on several mechanisms has been considered, including, the use of allosteric enzymes that have substrates with self-promoter properties,[94] "intensity filtering" (defined shortly) by redox transformations,[93] pH control by buffering,[80] and intensity filtering utilizing competing enzymatic processes.[80] These developments have built on earlier approaches to understand or realize sigmoid/digital (ON/OFF, YES/NO) responses in natural or synthetic biological and biochemical systems.[13,15,33,76]

The convex response, Figure 1(a), when scaled to the logic **0** to **1** input and output ranges, and assuming that the logic **0**s and at physical 0s (of the reactants' concentrations), always has slope larger than 1 at the origin, and therefore amplifies the spread of the input(s) due to noise, as it is transmitted to the output. In intensity filtering a fraction of the output[91,93,108] signal or that of the input signal(s) is neutralized[95] (or converted into one of the intermediate reagents) by an added chemical process, but only for small values of the signals. The two approaches are interrelated especially when the considered processes are networked: outputs then become inputs to other gates. The partial output removal approach has been successfully applied to systems of interest in applications,[42] as well as yielded realizations[41,124] of double-sigmoid (means, with "filtering" properties with respect to both inputs) AND and OR logic gates. As sketched in Figure 1(c), the price paid when using such "intensity filtering" is the potential loss of some of the signal intensity (and spread between the physical values corresponding to the binary **0** and **1**).

Intensity filtering based on partial input neutralization has been theoretically analyzed[30] for optimizing the binary output signal. In this section we survey this approach as an example. In the next subsection we describe the system for which experimental data were obtained in Ref. 95. We illustrate how a simplified kinetic description of the enzymatic processes involved can be set up, to limit the number of fitted parameters to key rate constants. Furthermore, ideally the model setup should be done in a way that allows us to identify those chemical or physical parameters of biocatalytic processes which could be adjusted to control the quality of the realized sigmoid response. Quality measures of the sigmoid response should then be optimized, including the steepness and symmetry of the sigmoid curve, as well as the issue of avoiding too much signal intensity loss due to the added filtering.

### 2.2. *Sigmoid Response Achieved by Neutralizing Some of the Input*

As an example, we analyze a specific system[95] that corresponds to signal transduction: The simplest "identity" logic gate that converts a single input: **0** or **1**, to the same binary value, **0** or **1**, of the output. In principle, the physical "logic values" (or ranges) of inputs that are designated as **0**s or **1**s are determined by the application. In fact, logic **0** needs not necessarily be at the physical zero. In the present case[95] the input is glucose in solution, the initial $t = 0$, where $t$ is the time, concentration of which can be varied. We take the experimental[95] input values 0 mM and 10 mM, for the binary **0** and **1**, respectively.

The signal processing was biocatalyzed by an electrode-immobilized enzyme glucose oxidase, resulting in oxidation of glucose. The output was measured[95] at the "gate time," $t_g = 180$ sec, as the current resulting from the transfer of two elementary units of charge per each oxidation cycle. In Figure 2, the current, $I(t_g)$, normalized per its maximum value $I_{max}(t_g)$ for the largest glucose input, $G_{max}$ = 10 mM, in plotted vs. the glucose input. The data are taken from Ref. 95, whereas the model fit, detailed later, is from Ref. 30.

For evaluating the effects of noise[72,91] in the signals, we have to consider the shape of the whole response curve, e.g., Figure 2, i.e., the output current vs. the input glucose concentration not only near the logic points but also generally over the whole **0** to **1** interval of values and somewhat beyond. As expected, the response curve here is convex. As described earlier, it is useful to convert the reponse to sigmoid, which offers advantages in noise handling, because small or zero slope at both logic points results in suppression of noise in the input as it is converted to the output.

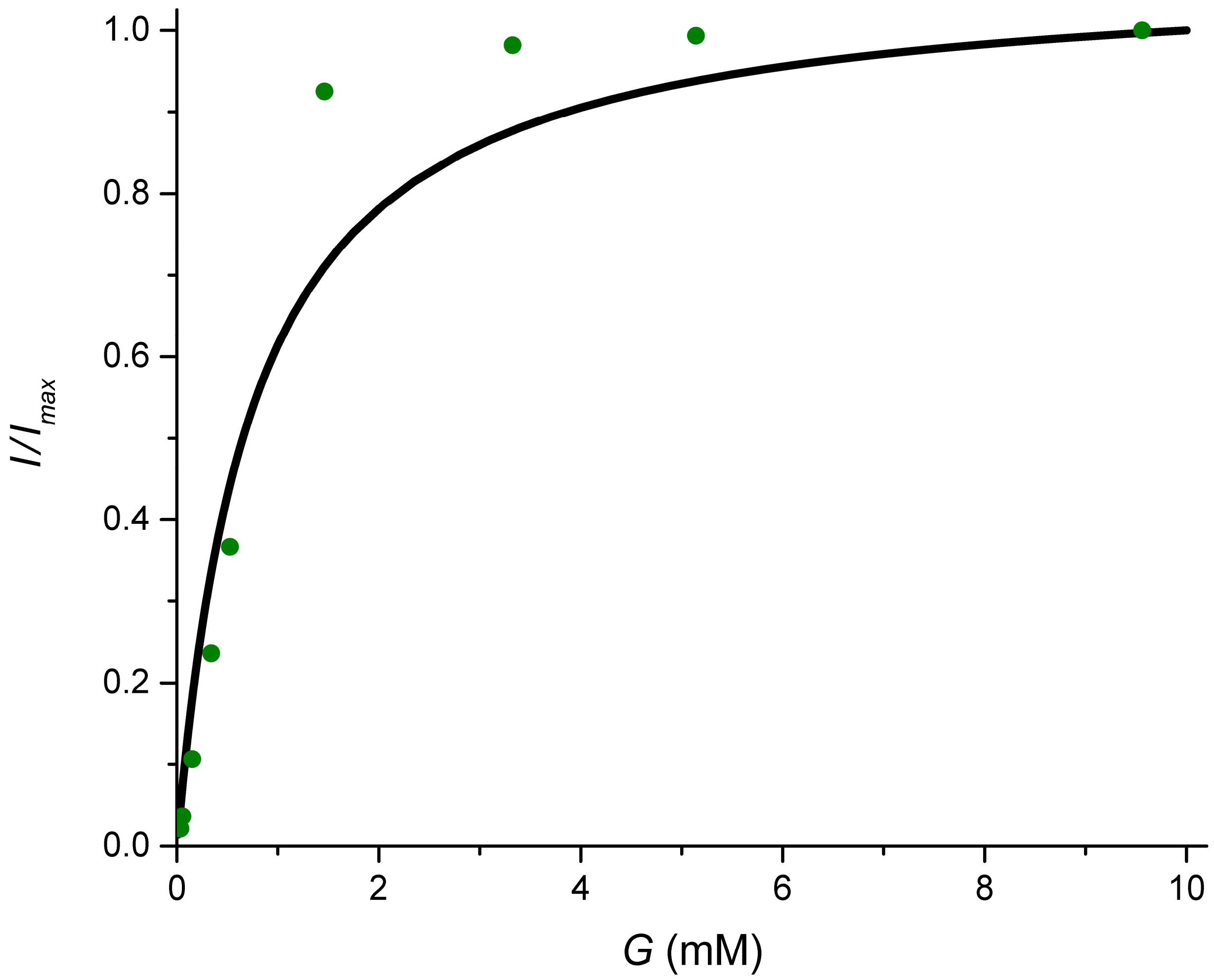

**Figure 2.** Experimental data[95] (circles) and our numerical model (line) for the normalized current at time $t_g$ vs. the *initial* glucose concentration, $G$, without the "filter" process, with the fitted parameters as described in the text. Adapted with permission from Ref. 30. Copyright © 2012, American Chemical Society.

Here we consider the approach[30] realized in Ref. 95, of neutralizing (consuming) a fraction of the input (glucose) in an added competing chemical process that only can use up a *limited amount* of glucose. Enzyme hexokinase was added to the solution, and adenosine triphosphate (ATP) was introduced in limited amounts as compared to the maximum 10 mM of glucose, to “switch on” the filtering effect. Indeed, the process biocatalyzed by hexokinase consumes glucose but only to the extent that ATP is not used up, without contributing to the output current. This makes the output signal, the current at the electrode, sigmoid. The corresponding experimental points from Ref. [95] and model fit (detailed shortly) are shown in Figure 3.

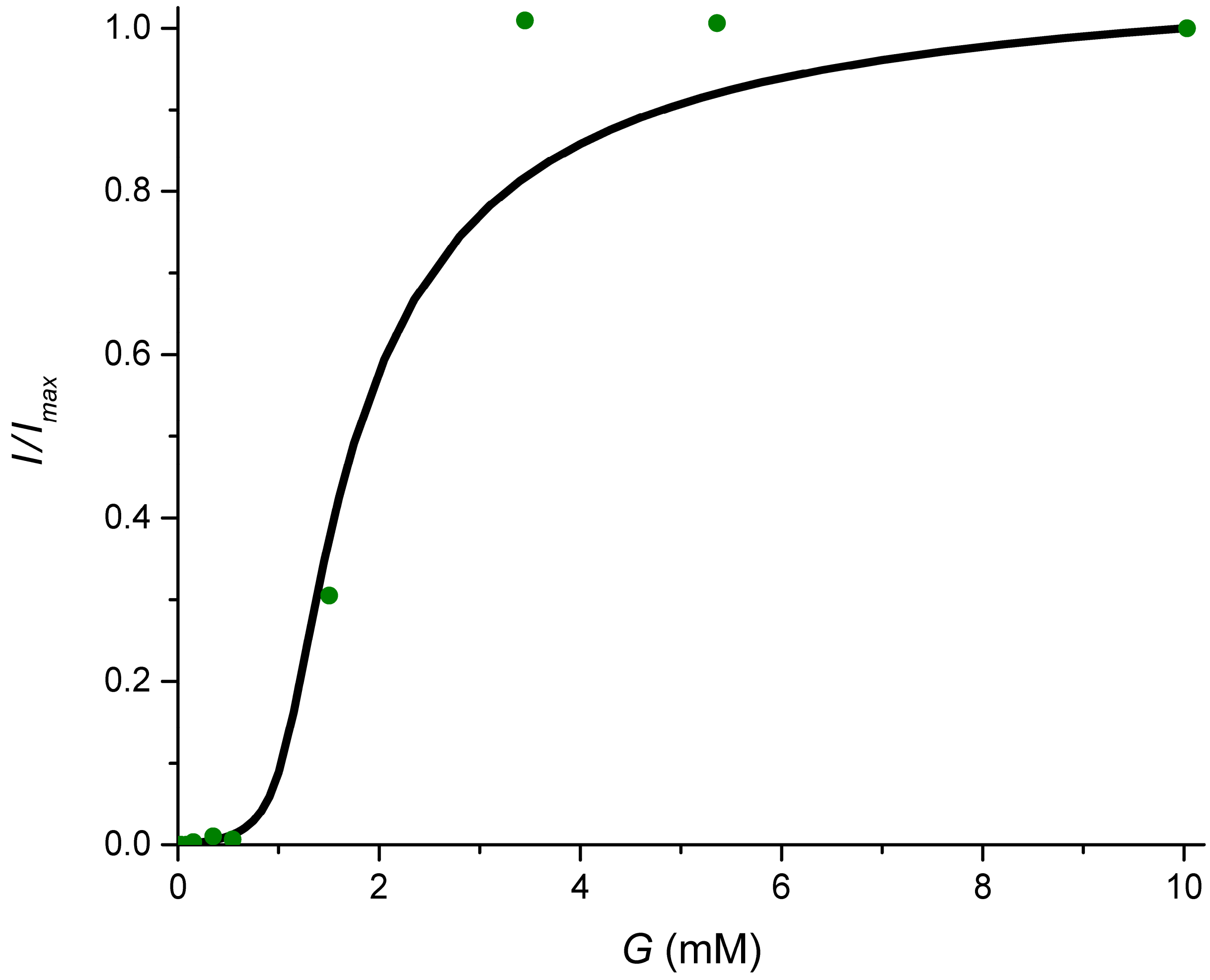

**Figure 3.** Experimental data[95] (circles) and our numerical model (line), the same as in Figure 2, but with the filtering process active. The process here is the same as in Figure 2, but with added hexokinase (2 μM). The initial concentration of ATP, 1.25 mM, was a fraction of the maximum initial glucose concentration, 10 mM. Adapted with permission from Ref. 30. Copyright © 2012, American Chemical Society.

As mentioned earlier, entirely phenomenological data fitting with properly shaped (convex or sigmoid) curves in not satisfactory, because we want to explicitly identify and model the dependence on those parameters which could be controlled to optimize the system's response. Phenomenological approaches[46] include the Hill function fitting.[95] Here we survey a different approach based on rate-equation modeling of the key steps of the enzymatic processes. We identify the concentrations of hexokinase and ATP as parameters to change, to significantly improve the sigmoid response.

Due to complexity of most enzymatic reactions, our modeling we focus on few key processes for each of them. Indeed, as emphasized earlier we want few adjustable parameters, suitable for the noisy data available in this field, e.g., Figures 2 and 3. With numerous parameters the specific data set might look better fitted, but the extrapolative power of the model will be lost. Thus, only enough adjustable parameters are kept to have a schematic overall-trend description of the response curves such as those shown in Figures 2 and 3.

Let us first consider glucose oxidase (GOx) only, without the added "filtering." We identify the following key process steps and their rates:

$$E + G \overset{k_1}{\rightarrow} C \overset{k_2}{\rightarrow} E + \cdots \, . \qquad (1)$$

Here $E$ denotes the concentration of GOx, and $G$ that of glucose. The intermediate products are produced in the first step, involving concentration $C$ of the modified enzyme. For glucose, unlike some other possible substrates for GOx, the first step (lumping several processes) can usually be assumed practically irreversible.[43-46] The last step is also irreversible. It is important to emphasize that we do not aim at a detailed kinetic study of the enzymatic reactions involved. As pointed out earlier, we seek a simple, few-parameter description of the response curve based on data from Ref. 95. We ignore the kinetics of all the other reactants, input or product, except for the rate equation for $C(t)$,

$$\frac{dC(t)}{dt} = k_1 G(t) E(t) - k_2 C(t) \, , \qquad (2)$$

which should be solved with $E(t) = E(0) - C(t)$. Indeed, we need this quantity only, because the current is proportional to the rate of the second step in Eq. (1),

$$I(t) \propto k_2 C(t) \,, \qquad (3)$$

i.e., our output is $I(t_g) \propto C(t_g)$.

Without the hexokinase "filtering" part of the process, we can assume that the GOx reaction at the electrode practically does not consume glucose: $G(t) = G(0) = G$. This assumption is appropriate for electrochemical designs for glucose sensing.[39,112,113] We also assume that the oxygen concentration is constant (and therefore is absorbed in a rate constant), ignoring the fact that for the largest glucose concentrations some corrections might possibly be needed due to oxygen depletion at the electrode.[95] With these assumptions, Eq. (2) can be solved in closed form,

$$C(t) = \frac{k_1 E(0) G}{k_1 G + k_2}\left[1 - e^{-(k_1 G + k_2)t}\right] . \qquad (4)$$

Here the initial (and later remaining constant) value of $G$ is the input, varying from 0 to $G_{max}$. For logic-gate-functioning analysis of such processes, we define scaled logic-range variables,

$$x = \frac{G(0)}{G_{max}} \,, \quad y = \frac{I\big(t_g; G(0)\big)}{I_{max}} = \frac{C\big(t_g; G(0)\big)}{C\big(t_g; G_{max}\big)} \,, \qquad (5)$$

where $I_{max} = I\big(t_g; G_{max}\big)$. The slope of $y(x)$ near the logic point values $x = 0$ and 1, determines the noise transmission factors.[86,91]

The data in Ref. 95 were given as the values of $y$ for several inputs, $G(0)$. Without the filter process, least-squares fit of these data in our case yielded the estimates $k_1 \cong 80 \text{ mM}^{-1}\text{s}^{-1}$, $k_2 \cong 60 \text{ s}^{-1}$. However, these estimates are rather imprecise, as indicated by the numerical fitting procedures. Indeed, these rate constants are large in the sense that the dimensionless combinations $k_2 t_g$ and $k_1 G_{max} t_g$ are both much larger than 1. This is consistent with other estimates of these rate constants for GOx with glucose as a substrate.[12,31,38,61] The dependence of the scaled variable $y$ on $G = G(0)$,

$$y = \frac{\frac{k_1 E(0) G}{k_1 G + k_2}\left[1 - e^{-(k_1 G + k_2) t_g}\right]}{\frac{k_1 E(0) G_{max}}{k_1 G_{max} + k_2}\left[1 - e^{-(k_1 G_{max} + k_2) t_g}\right]} \approx \frac{G\left(G_{max} + \frac{k_2}{k_1}\right)}{G_{max}\left(G + \frac{k_2}{k_1}\right)} , \tag{6}$$

is then to a good approximation only controlled by the ratio $k_2/k_1$, for which a relatively precise estimate is possible, $k_2/k_1 = 0.75 \pm 0.02$ mM. The quality of the fits such as that shown in Figure 2, is not impressive, but this is similar to the situation with the more phenomenological Hill-function fitting.[95]

With the filter process added, in the presence of hexokinase (HK), of concentration denoted $H(t)$, and ATP, of concentration $A(t)$, glucose will be depleted. Data are then available[95] for several initial values $A(0)$, all smaller than $G_{max}$. In order to limit the number of adjustable parameters we will only consider that pathway of the HK biocatalytic process[36,118] in which glucose is taken in as the first substrate, to form an intermediate product of concentration $D(t)$. We again take a simplified scheme for the HK activity, ignoring a possible reversibility of the complex formation and other details,[36,37,45,118]

$$H + G \overset{k_3}{\rightarrow} D + \cdots , \quad D + A \overset{k_4}{\rightarrow} H + \cdots . \tag{7}$$

This approach yields only two adjustable parameters which enter the rate equations that determine the time-dependence of glucose to use in Eq. (2) for calculating $C(t)$,

$$\begin{aligned} \frac{dG}{dt} &= -k_3 H G , \\ \frac{dH}{dt} &= -k_3 H G + k_4 D A , \\ \frac{dD}{dt} &= k_3 H G - k_4 D A , \\ \frac{dA}{dt} &= -k_4 D A . \end{aligned} \tag{8}$$

Note that the two middle equations can be made into one by using $D(t) + H(t) = H(0)$. The available data were for $H(0) = 2$ μM. The resulting system was solved numerically, and the data available for the four initial nonzero ATP concentrations were fitted to yield the estimates $k_3 = 14.3 \pm 0.7 \mathrm{mM^{-1}s^{-1}}$, $k_4 = 8.1 \pm 0.4\ \mathrm{mM^{-1}s^{-1}}$. The earlier estimate for $k_2/k_1$ was used to obtain these values.

### 2.3. *Sigmoid Response Optimization*

For fault-tolerant[3,32,90] information processing when gates are connected in a network,[35,116] parameters must be chosen to reduce the analog noise amplification or avoid it, the latter accomplished by filtering. There are various sources of noise in the biochemical reaction processes that affect their performance as binary "gates." Imprecise and/or noisy realization of the expected response curve, $y(x)$, is one such source. There is also noise in the input(s) that is passed to the output. In biochemical environments the noise in the inputs is quite large.[23,26,51,52,86,91,97,110] Avoiding this "analog noise" being amplified during signal processing is paramount to small-scale network stability. For larger networks, additional consideration of "digital" errors[86] is required, but here we focus on the single gate design.

Unless the input noise levels are very large or the response curve has non-smooth features near the logic point $x = 0$ or 1, then the noise transmission factor is simply given by the slope of the curve $y(x)$ near each of the two logic points. Filtering can make both slopes (at **0** and **1**) much smaller than 1, compare Figures 2 and 3. For best results, the filtering response-curve shape should be centered away from **0** or **1** and also steep. However, improvement of the quality of filtering should not be done at the expense of preserving the intensity of the output signal in terms of its actual range of values, here equal $I_{max}$, as opposed to the scaled variable $y$. Loss of intensity amplifies the *relative level* of noise from all the sources discussed above.

The inputs are set by the gate usage and typically cannot be adjusted. We can select other parameters values to optimize the filtering quality. Here we formulate quantitative criteria for such optimization. Note that within the assumed regime of functioning, in our model the shape of $y(x)$ does not depend on $E(0)$. However, other "gate machinery" (means, not input or output) initial chemical concentrations can be varied. Here we consider the adjustment of $H(0)$ and $A(0)$. Other modifications can include changing the physical or chemical conditions (which affects the rates of various processes) or limiting the supply of oxygen.[71]

To have the response curve as symmetric as possible we consider the position of the peak of the slope, $y'(x)$. In enzymatic processes, sigmoid response-curves are typically not symmetrical with respect to the inflection region; see Figure 3 and also some approximate analytic expressions and their plots in Refs. 89, 92. We can define the width of the peak of the derivative by the difference $x_2 - x_1$, where by $y'(x_{1,2}) = 1$. The middle-point of the peak is defined at $(x_2 + x_1)/2$. Figure 4 shows three different illustrative sigmoid response curves, as well as their derivatives calculated in our model, with the parameter values discussed in the preceding subsection. Figure 5 presents a contour plot of the *deviation* of the middle-point peak position from 1/2. Our aim is to get it rather close to 1/2 without compromising the other gate-quality criteria. One of these is analyzed in Figure 6, which plots the width of the peak, which we would like to be as small as possible.

A "non-binary" criterion for gate quality is that of avoiding to the extent possible the loss of the signal intensity. Since enzymatic processes usually approach saturation at large inputs, here this can be defined as the fractional loss:

$$1 - \frac{I_{max}(H(0)>0, A(0)>0)}{I_{max}(H(0)=0, A(0)=0)} = 1 - \frac{I(t_g, G_{max})_{H(0)>0, A(0)>0}}{I(t_g, G_{max})_{H(0)=0, A(0)=0}} . \tag{9}$$

This quantity is shown in Figure 7 as the percentage value. Figures 5, 6, 7 span values safely within the experimentally realizable ranges of the considered control quantities, $H(0)$ and $A(0)$. Consideration of Figures 5 and 6 suggests that the peak can be made optimally centrally positioned and narrow, by selecting approximately $H(0) = 4\ \mu\text{M}$ and $A(0) = 4\ \text{mM}$. The optimal choices correspond to the regions marked by the white ovals in the figures. At least some loss of intensity is usually present for this type of filtering. However, it can be tolerated if percentage-wise it is comparable to (or smaller than) the degree of noise otherwise present in the output. The approximately 5% loss level in the oval-delineated region (see Figure 7) is therefore acceptable.

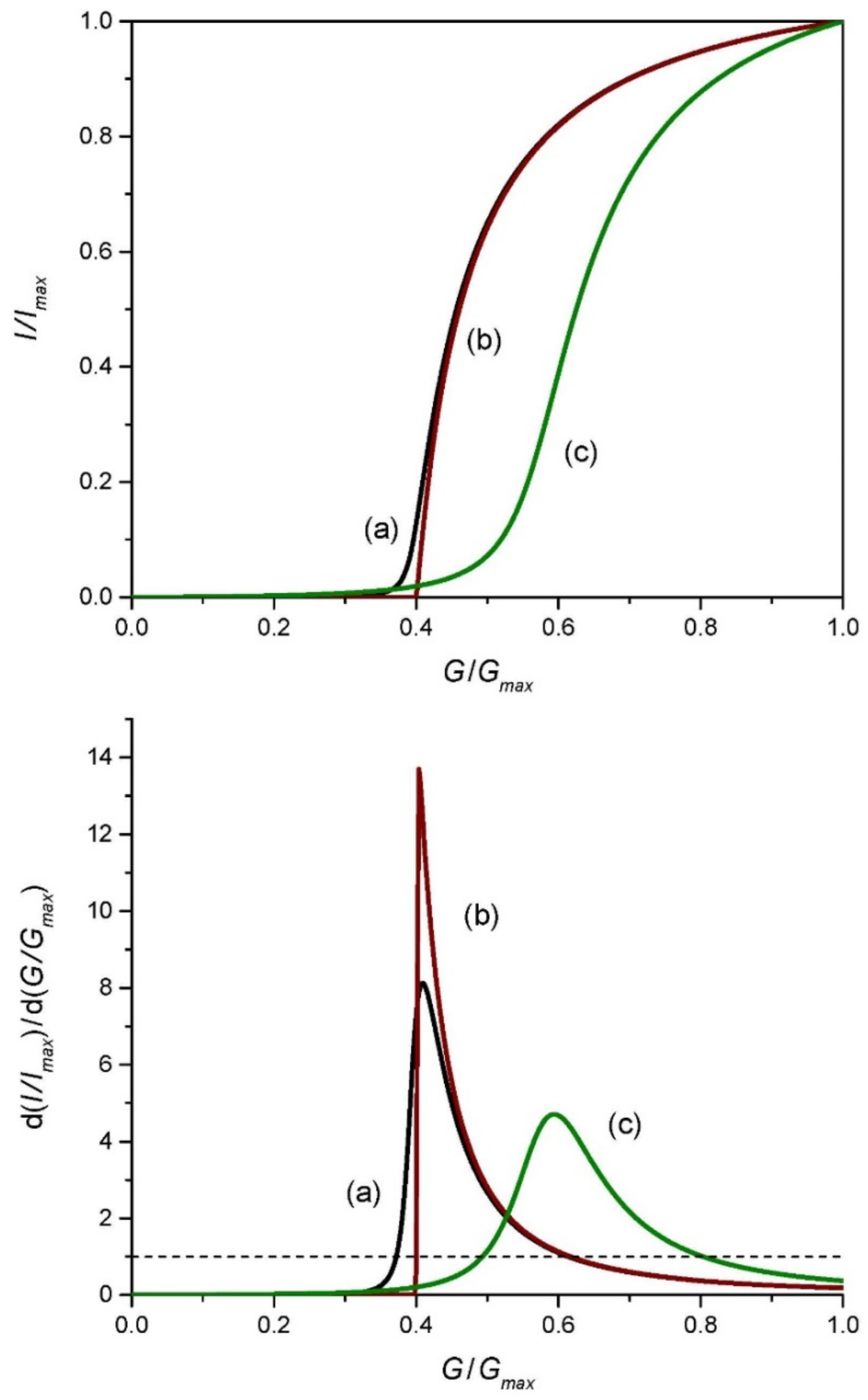

**Figure 4.** Examples of sigmoid curves (top panel) and their derivatives (bottom panel) for three different selections of the parameters used to control the response: (a) $H(0) = 4$ µM and $A(0) = 4$ mM; (b) $H(0) = 8$ µM and $A(0) = 4$ mM; (c) $H(0) = 3$ µM and $A(0) = 6$ mM. The values (a) correspond to the center of the optimal range as described in the text. The dashed line indicates the level at which the width of the peak of the derivative is measured. Adapted with permission from Ref. 30. Copyright © 2012, American Chemical Society.

Our optimal sigmoid response shape and its derivative are shown as curves (a) in Figure 4. While not symmetrical, the response curve is centrally positioned and rather narrow. The derivative of the output signal in regions $0 \leq x \lesssim 0.37$ and $0.63 \lesssim x \leq 1$ is less than 1, see Figure 4: bottom panel, curve (a). In these two regions, each extending ~ 37% from the logic points **0** and **1**, on the input axes, the noise in the input will not be amplified. The criteria just surveyed are quite general and can be applied to other systems contemplated for information and signal processing or for biosensing with biomolecular processes.

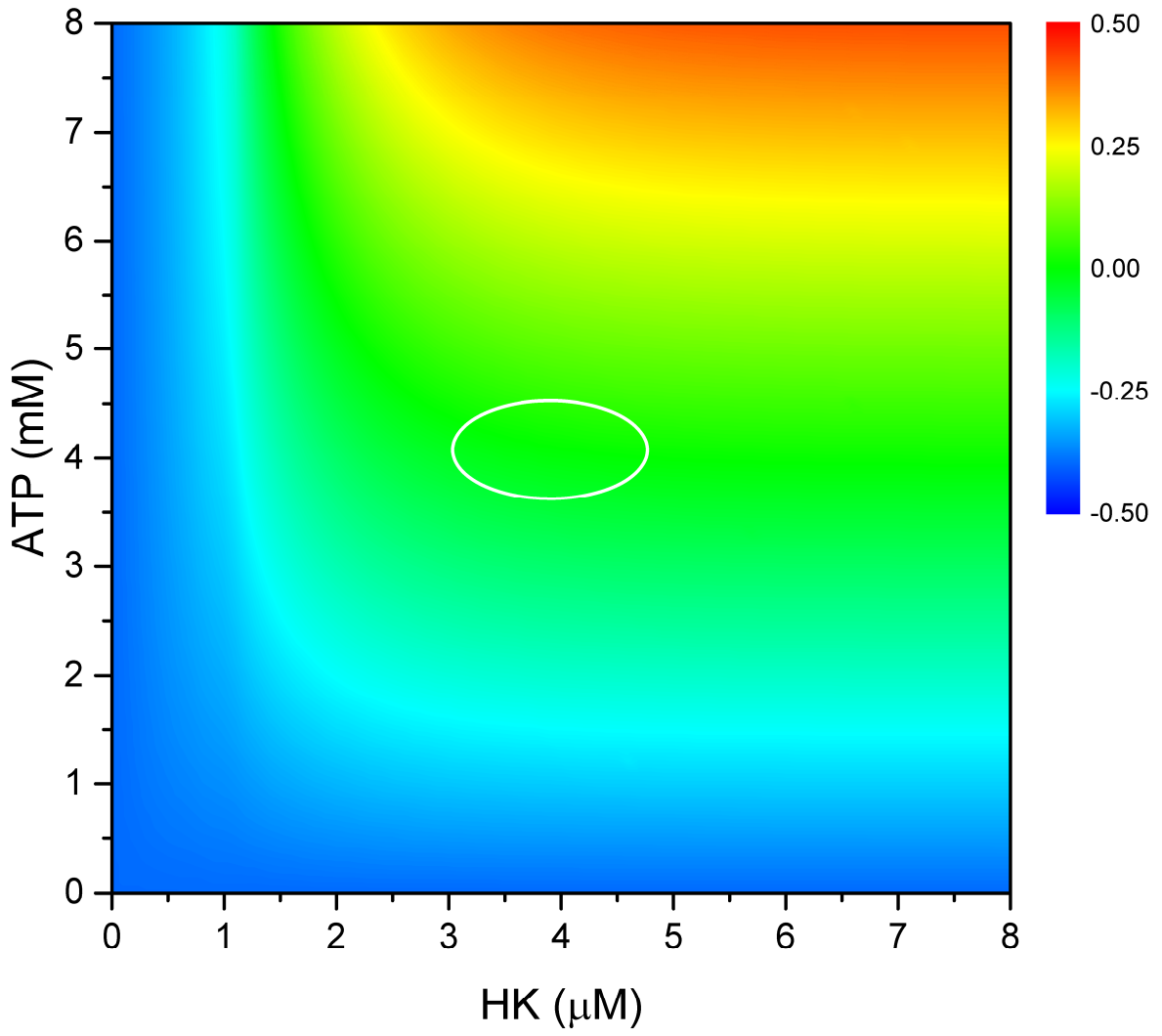

**Figure 5.** Contour plot for various initial values of HK and ATP, of the deviation of the middle-point of the peak location from 1/2, i.e., $(x_2 + x_1 - 1)/2$. The optimal values are as small as possible (green color). The oval defines the best choice of the parameters considering the other criteria for optimizing the response: see text. Reprinted with permission from Ref. 30. Copyright © 2012, American Chemical Society.

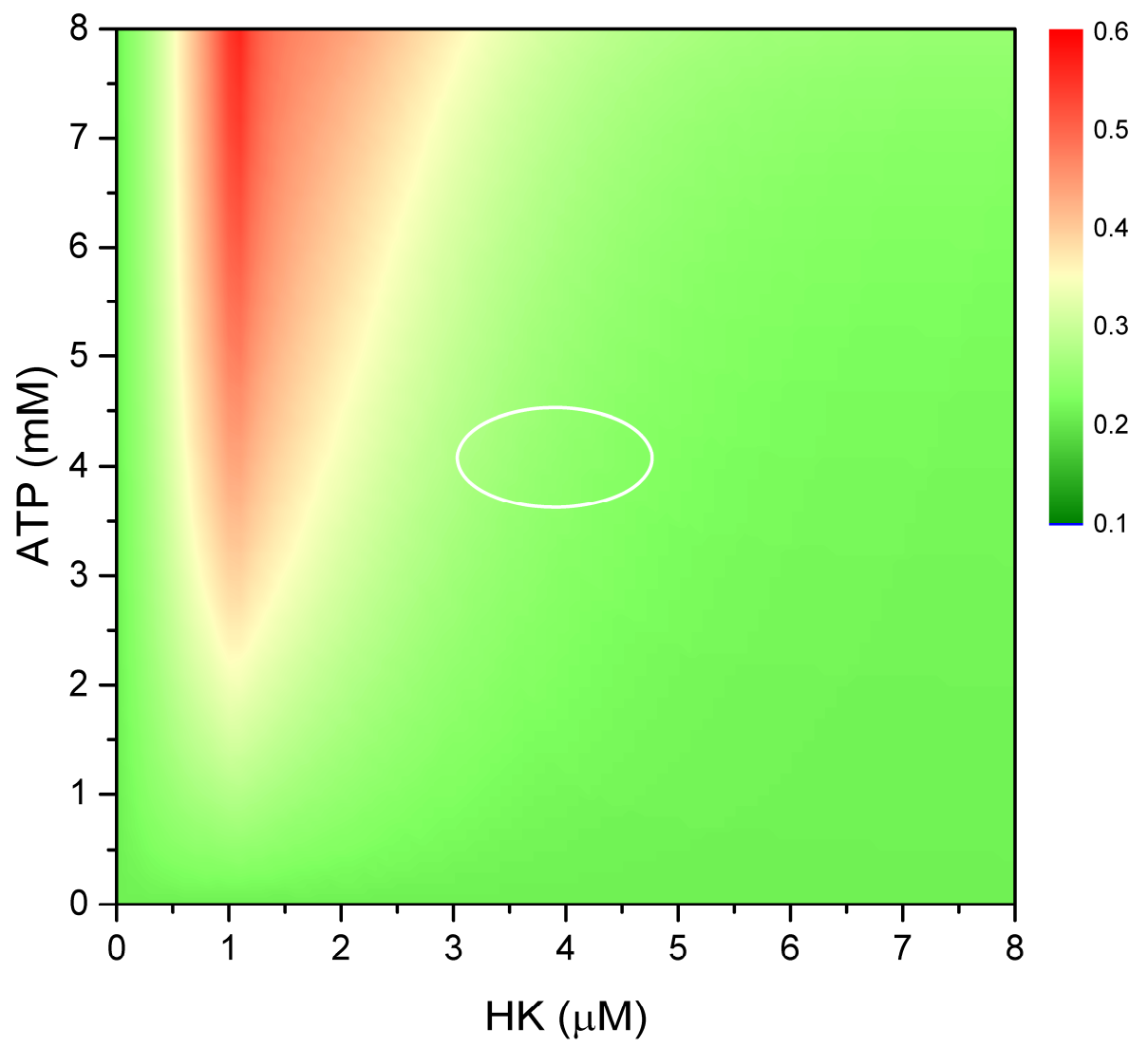

**Figure 6.** Contour plot of the width of the peak. The optimal values are as small as possible (the green shades). The oval defines the best choice of the parameters considering the other criteria for optimizing the response: see text. Reprinted with permission from Ref. 30. Copyright © 2012, American Chemical Society.

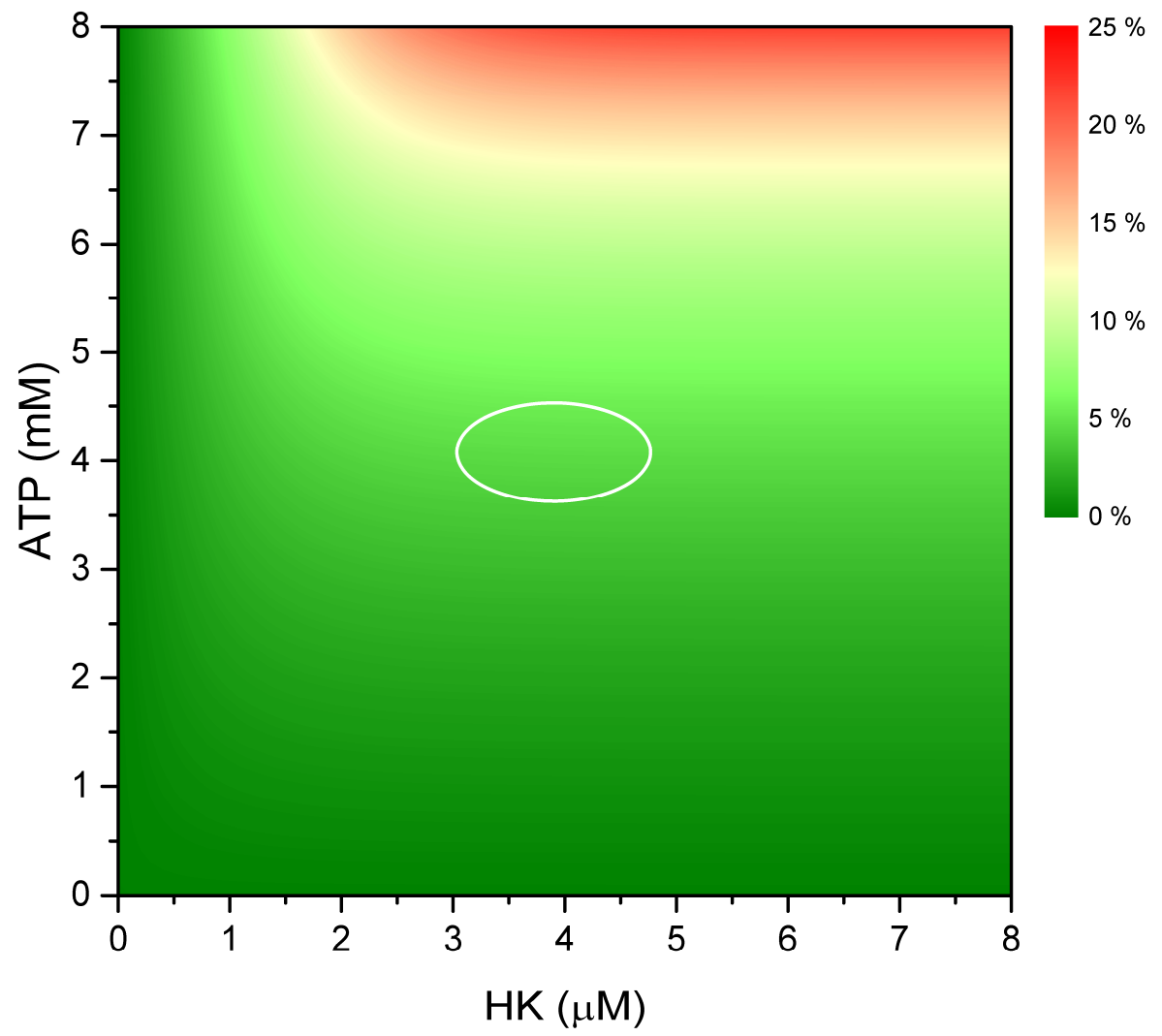

**Figure 7.** Contour plot of the measure of the loss of the output signal intensity, Eq. (9). This measure should be minimized (green color) without compromising the other gate-quality criteria. The oval defines the best choice of the parameters considering the other two criteria: see text. Reprinted with permission from Ref. 30. Copyright © 2012, American Chemical Society.

## 3. Threshold Response in an Enzymatic System

### 3.1. *Modifications of Response Functions of Biomolecular Processes*

In the preceding section we considered conversion of convex response to sigmoid. However, in various applications different changes in the response function might be desirable. In biosensing applications in many situations it is useful to modify the generic response to make it as linear as possible,[5,20,22,25,29,34,75,87,96,101-103,111,122,128] i.e., the conversion (a) ➡ (b) in Figure 1, here also hoping to avoid too much loss in the overall signal intensity. Recently, a model was developed[87,122] (not reviewed here) and applied to data analysis, of how two enzymatic processes with different nonlinear responses can be combined to yield an extended approximately linear response regime.

Recently, experiments[66] on three-input majority and minority enzymatic gates for biocomputing applications have underscored the importance of another type of "biochemical filtering" as a part of the biochemical post-processing of the output to achieve the desired response. In this case the conversion of a linear response to the threshold one, followed by a linear behavior, (b) ➡ (d) in Figure 1, is required. Here we review results[88] establishing that such "filtering" mechanism in the reported experiments[66] (and in the earlier work on filtering[67]) utilizing the enzyme malate dehydrogenase (MDH), also called malic dehydrogenase, is a result on an unusual mechanism of enzymatic biocatalytic activity of this enzyme, noted in an early work on the mechanism of action of MDH.[100] This work[100] considered what is called[69] a reversible random-sequential bi bi mechanism of action for MDH, and reported that MDH can undergo a variant of inhibition[100] that results in the slowing-down of the oxidation/reduction of one of the two substrate/product redox couples.

As suggested by this observation, modeling of the filtering effect here is quite different from that for the afore-surveyed[6,30,41,92,122-124] "intensity filtering," We survey an appropriate description, which was applied[88] to data for a system where the initial linear response is due to the biocatalytic action of another enzyme, glucose dehydrogenase (GDH). We also report (in the next subsection) additional interesting conclusions for "intensity filtering" that was considered in the preceding section.

### 3.2. *Signal Transduction Combined with Fast Reversible Deactivation of the Output*

The system that is considered here is shown schematically in Figure 8. We already emphasized that the full kinetic description of enzymatic processes requires several rate constants for each enzyme. We will revisit this later (in Section 3.3). Let us first attempt to use a simple model with a minimal number of parameters in an attempt to describe the effect on a linear response of the type shown in Figure 1(b), of an added process that affects the output product, of concentration, $P(t)$, by rapidly interconverting it to and from (equilibrating it with) another compound that is inert as far as contributing to the output signal. Our conclusion will be that this simple description is not adequate for the system of interest.[88] However, the model itself is useful to study because adding fast, reversible processes that affect the product can be done relatively easily in most cases by chemical or biochemical means.

The first enzyme in the cascade, GDH, was utilized in the kinetic regime quite typical for many uses of enzymes, i.e., with both of its input chemicals (substrates), glucose and $NAD^+$, provided with the initial concentrations large enough to have the products of the reaction generated with a practically constant rate for the times of the experiment. For the product of interest, NADH, we thus assume that its concentration, $P(t)$, varies according to

$$\frac{dP}{dt} = RG, \qquad (10)$$

$$P(t_g) = RGt_g, \qquad (11)$$

where $R$ is a rate constant that can be fitted from the data, whereas $G$ is the initial concentration of glucose, which is the input at time $t = 0$, was varied from 0 up to, here, $G_{max} = 8$ mM. Other reagents in the present system have fixed initial concentrations. The linear behavior in time applies for all but the smallest inputs, $G$, and it breaks down for very short times as well as for very long times on the time-scales of the experiments that went up to 600 s.

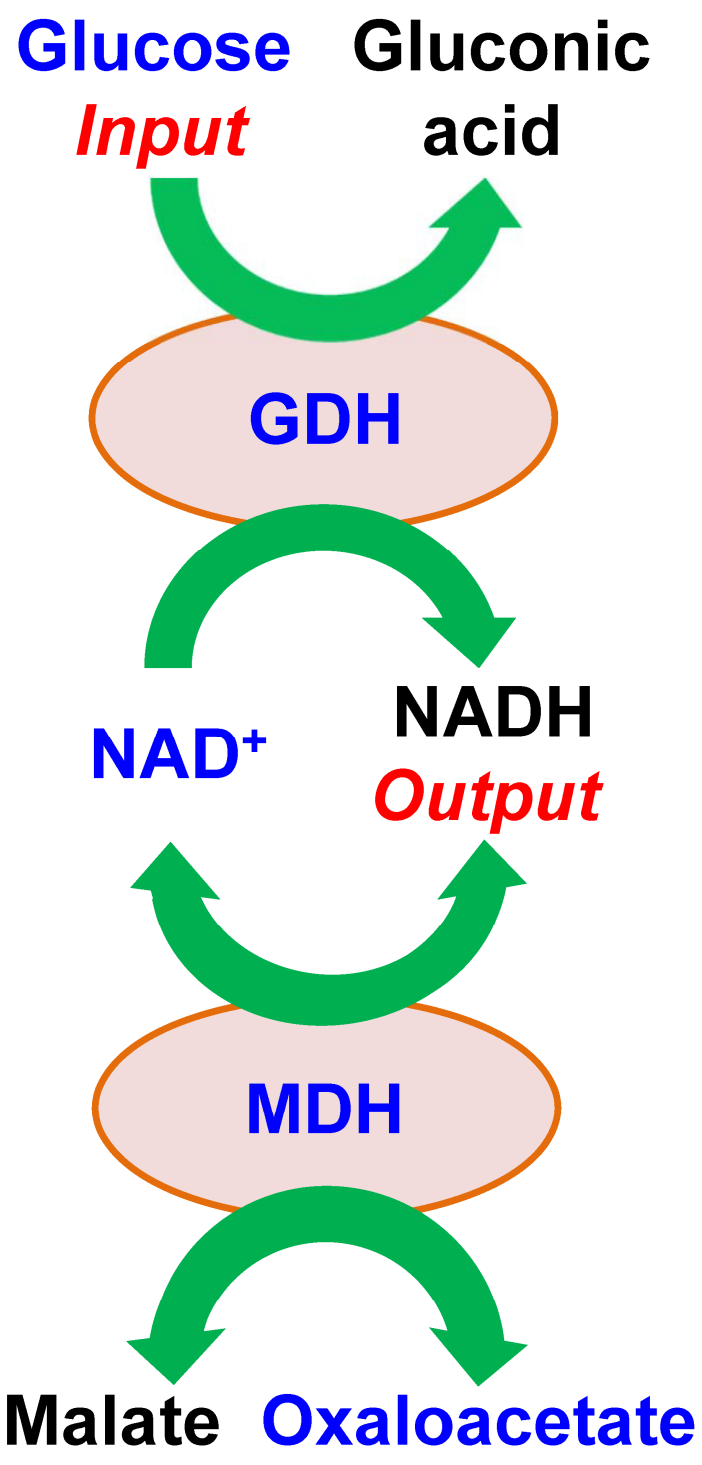

**Figure 8.** The schematics of the enzymatic processes in the biocatalytic cascade[88] surveyed in Section 3. The reactants and biocatalysts that are initially in the system (with filtering) are color-coded blue, including β-nicotinamide adenine dinucleotide ($NAD^+$) and its reduced form (NADH). The double-arrows emphasize that the functioning of MDH is reversible. Reprinted with permission from Ref. 88. Copyright © 2014, American Chemical Society.

The second enzyme, MDH, is also used in the regime of plentiful supply of the initially available substrates (one of the two in each direction of functioning, see Figure 8). Since its functioning is reversible, we could attempt to describe the kinetics of the present system by the effective processes

$$G \xrightarrow{R} P, \qquad P \underset{r_-}{\overset{r_+}{\rightleftarrows}} M. \qquad (12)$$

We note that MDH oxidizes NADH to $NAD^+$, which is then our "inert" compound (not contributing to the measured signal obtained by optically detecting the concentration of NADH), but since $NAD^+$ is already present in the system in a large quantity, the variation of its concentration has little effect on the reverse process. However, malate, denoted, $M(t)$, see Figure 8, not initially present, directly (and for simplicity we assume linearly) affects the reverse process rate. The present model is not accurate, but interesting because the resulting rate equations can be solved in closed form,

$$\frac{dP}{dt} = RG - r_+P + r_-M, \qquad \frac{dM}{dt} = r_+P - r_-M, \tag{13}$$

$$P(t) = RG\left\{\frac{r_+\left[1-e^{-(r_++r_-)t}\right]}{(r_++r_-)^2} + \frac{r_-t}{r_++r_-}\right\}. \tag{14}$$

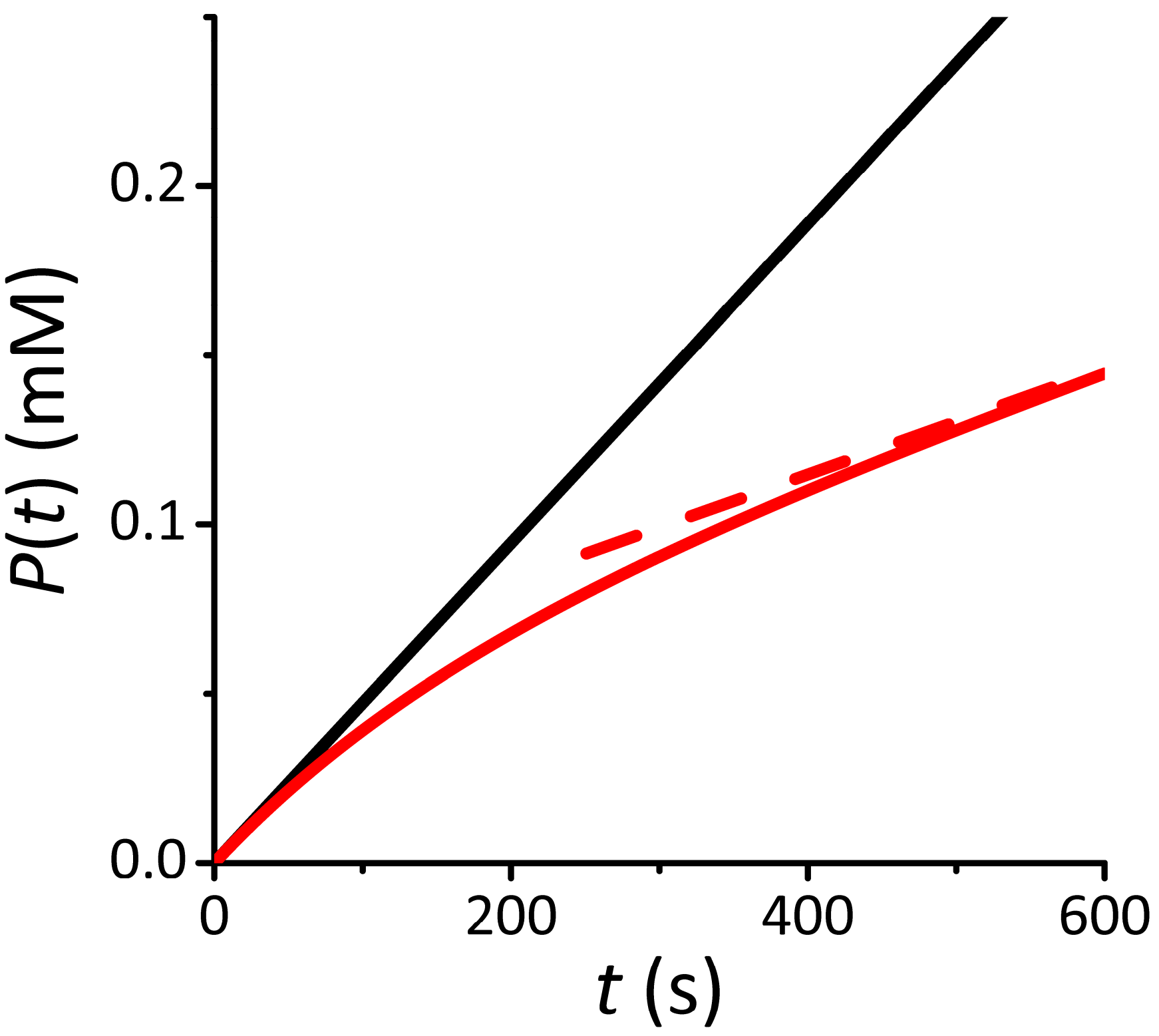

**Figure 9.** Time dependence of the NADH concentration, $P(t)$, for typical parameter values[88] with (the red curve) and without (the black straight line) the added fast reversible "output deactivation" process biocatalyzed by MDH. The dashed line is the asymptotic rate, $RGr_-/(r_+ + r_-)$, for large times. Adapted with permission from Ref. 88. Copyright © 2014, American Chemical Society.

One could speculate that an added fast reversible process that deactivates a part of the product, up to a fraction that corresponds to the concentrations of the rate-limiting chemicals for which that reversible process equilibrates, might have some "filtering" effect. But Eq. (14) suggests that there is no "filtering" at all. Instead, the dependence of the product $P(t_g)$ on the input, $G$, remains linear for any fixed "gate time" $t_g$, with a reduced slope (means, with loss of intensity). The original time-dependence, Eq. (11), is linear in both $G$ and $t_g$. However, with the added process the time dependence is modified. Figure 9 illustrates that for small times the rate of the product output is unchanged (the added process is not really active). For large times a reduced rate, $RGr_-/(r_+ + r_-)$, is approached.

Interestingly, the experimentally observed[66] change from the linear to threshold response, (b) ➡ (d) in Figure 1, must therefore be due to more complicated kinetic mechanisms than that summarized in Eq. (12). The origin of the observed effect turns out to be connected to an interesting kinetic property of the functioning of MDH, reviewed in the rest of this section. The model just considered, however, suggests that, generally adding a fast, reversible process of deactivation of the input by *equilibrating* it with another species cannot in itself result in threshold type (at low inputs) intensity filtering. Examples[6,30,41,42,50,80,89,92,93,95,114,122-124] when such an approach worked have always involved the *absence of equilibration* by kinetic restrictions, for example due to a limitation on how much of the other species could be produced (imposed by the process requiring some other, limited-supply chemical).

### 3.3. *MDH Kinetics with Inhibition*

Enzymes have rather complicated kinetic mechanisms. These typically involve the formation of complexes with substrate(s), then follow-up processes involving these complexes, etc., in most cases resulting in the restoration of the enzyme at the end of the cascade, when products are released. Our first enzyme, GDH, has such a standard mechanism of action,[10,77,109] that would require several rate constants to fully model. The second enzyme, MDH, has a complicated and less common mechanism of action,[24,74,100,119] with a number of intermediate complexes. It is in fact not fully studied. MDH can form complexes[100] with all four of the relevant substrates for the direct (NADH and oxaloacetate) or reverse ($NAD^+$ and malate) functioning, and then form triple-complexes in which the actual redox-pair conversions occur. This is sketched in Figure 10(a). Modeling[18] of all the processes would require at least 18 rate constants. This illustrates why it is so important to use few-parameter kinetic models for a semi-qualitative description of the response in sensor and biomolecular computing applications. Such approaches[86,92] usually involve setting up an effective rate equation description that captures the main process pathways.

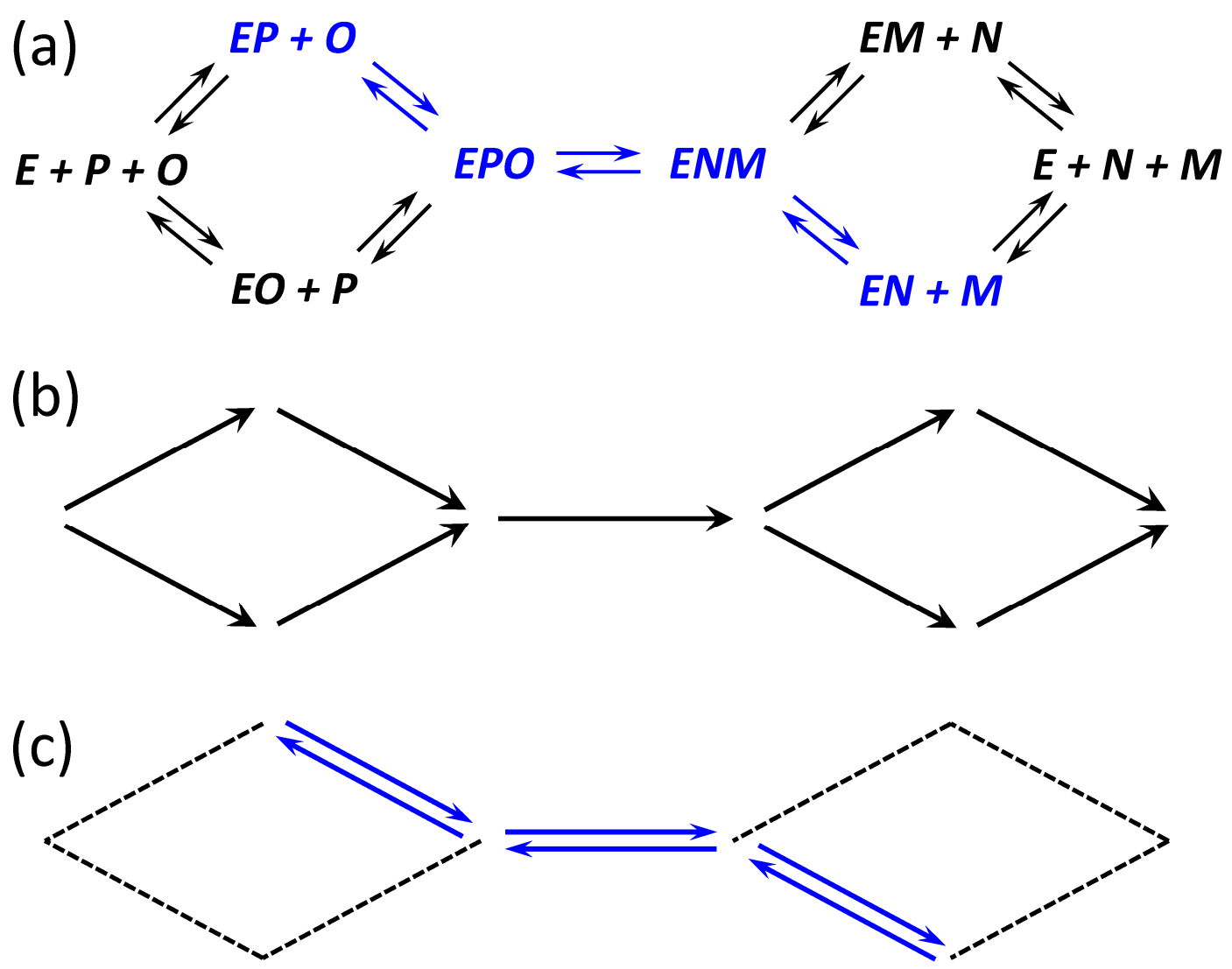

**Figure 10.** (a) Mechanism of action of MDH. Here *E* stands for the enzyme, *P* for NADH (the product), *N* for $NAD^+$, whereas malate and oxaloacetate are denoted by *M* and *O*, respectively. (b) The "direct" reaction pathways activate at early times. (c) A hypothetical mechanism for a reaction pathway subset that dominates at later times.

The output product, NADH, denoted *P*, generated by the GDH process, activates all the "direct" complex-formation and redox conversion processes of MDH, Figure 10(b). The latter not only partially convert NADH back to $NAD^+$, to be denoted *N*, but these processes also build up the concentration of malate, *M*. The "reverse" processes of MDH are them also activated, driving the system towards equilibration. However, it has been have reported in the literature[100] that, as the concentration of malate is increased relative to oxaloacetate, denoted *O*, the redox inter-conversion rate NADH ↔ $NAD^+$ actually slows down, whereas the inter-conversion rate oxaloacetate ↔ malate increases. This might look paradoxical, but a likely explanation is as shown in Figure 10(c). Most of the enzyme, *E*, becomes trapped in the complexes *EP* and *EN* (as well as in complexes, *ENM* and *EPO*). The fast inter-conversion oxaloacetate ↔ malate (*O* ↔ *M*) is compensated for by the inter-conversion *EP* ↔ *EN*. This interesting mechanism can be either kinetic or can caused by malate inhibiting[100] some of the reaction pathways. It is important to emphasize that despite the earlier experimental evidence,[100] this mechanism is largely a conjecture. In fact, the observation that this assumption leads to modeling[88] that fits the data provides an additional support to it.

To model this effect with a minimal possible number of parameters, considering that oxaloacetate is supplied in large quantity, we ignore its depletion. We assume that the concentration of malate that would correspond to steady state is $M_0$. We then write the rate equation of the linear supply of the product, cf. Eq. (10), but with the added depletion term,

$$\frac{dP}{dt} = RG - K(M_0 - M)P = -KP^2 - K(M_0 - Rt)P + RG. \qquad (15)$$

Here $K$ is the rate constant for the decrease in the amount of the product, *P*, due to the initially active mechanism, Figure 10(b), which is gradually replaced by the mechanism involving *EP* ↔ *EN* as $M$ increases from 0 to $M_0$, Figure 10(c). This assumes that the relative rates of the two mechanisms are directly proportional to $M_0 - M$ and $M$, respectively. The second expression in Eq. (15) was obtained by using $M(t) = RGt - P(t)$. This can be solved to yield

$$P(t) = RGt - M_0 + \frac{M_0 e^{-K\left(\frac{1}{2}RGt - M_0\right)t}}{1 + KM_0 \int_0^t e^{-K\left(\frac{1}{2}RG\tau - M_0\right)\tau} d\tau}, \qquad (16)$$

or

$$P(t) = RGt - M_0 + \frac{2\sqrt{KRG}\, M_0 e^{\frac{Kt}{2}(2M_0 - RGt)}}{\sqrt{2\pi} K M_0 e^{\frac{KM_0^2}{2RG}} \left[\operatorname{erf}\left(\sqrt{\frac{K}{2RG}} M_0\right) - \operatorname{erf}\left(\sqrt{\frac{K}{2RG}} (M_0 - RGt)\right)\right] + 2\sqrt{KRG}}. \qquad (17)$$

This expression provides the dependence of $P(t_g)$ on $G$, of the type shown in Figure 1(d), and was successful in experimental data fitting[88] for a system the schematic of which is sketched in Figure 8. Figure 11 provides an illustration of fitting the experimentally measured[88] time dependence, and also shows an example of data fitting[88] for the response function, which should be compared with Figures 1(b) vs. 1(d).

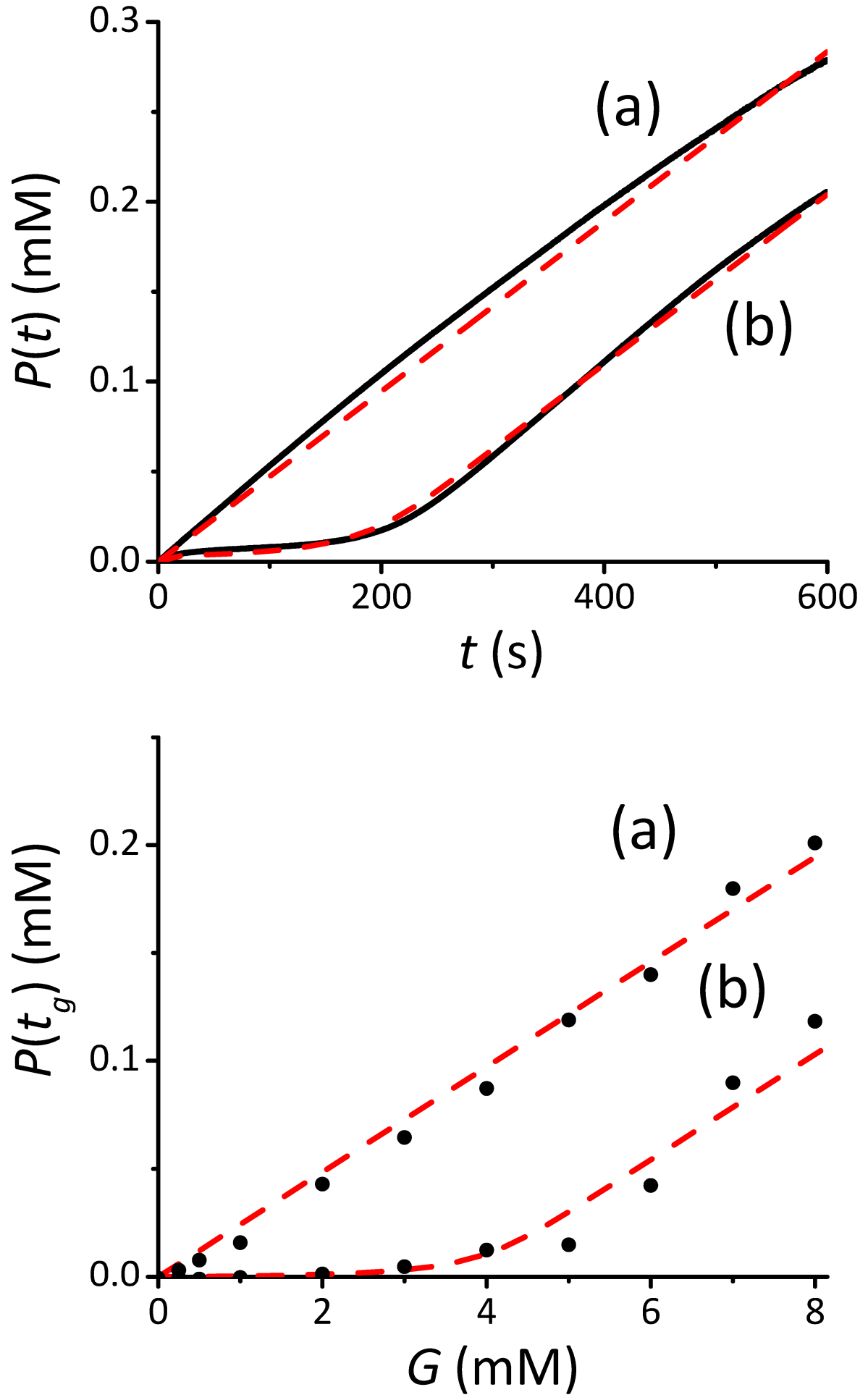

**Figure 11.** Top panel: Measured time dependence[88] for input $G$ = 7 mM, points practically merged into solid lines, and model results, shown as dashed lines, for a typical experiment (a) without filtering, and (b) with filtering. Bottom panel: Measured dependence on the initial glucose concentration[88] for fixed time $t_g$ = 360 s, shown as dots, for (a) without filtering, and (b) with filtering; model results are shown as dashed lines. Adapted with permission from Ref. 88. Copyright © 2014, American Chemical Society.

## 4. Summary

We reviewed the biochemical "intensity filtering," by considering approaches to modeling binary AND gate performance and optimization of its "digital" response. Specifically, we considered the recently introduced approach of a partial input conversion into inactive compounds, which yields sigmoid response of the output, of interest in information/signal processing and in biosensing applications. For selected examples, we established criteria for optimizing such a "binary" response. Different physical or chemical conditions can be changed to impact enzymatic processes, and we demonstrated this by an example of how our system's response changed when the initial concentrations of two "filter process" chemicals were varied. The developed criteria are quite general and can be applied to other systems contemplated for information/signal processing, and for biosensing, with biomolecular processes.

Applying a similar rate-equation modelling approach we then demonstrated that reversible conversion of the product to another compound cannot on its own result in (bio)chemical "filtering." Experimentally observed biochemical "threshold filtering" by a reaction biocatalyzed by an enzyme with an unusual mechanism of action was instead attributed to inhibition of certain process pathways for this enzyme once one of its substrates builds up in concentration. Experimental data analysis supports the model's validity.

## 5. Acknowledgements

The authors thank their many colleagues,[3,6,30,32,41,42,53,57-59,63,72,80,83-94,122-124] notably, Prof. E. Katz, for collaborative teamwork. They also gratefully acknowledge funding of the research work reviewed here by the US National Science Foundation, most recently under award CBET-1403208.